\documentclass[paper]{JHEP3}
\usepackage{amsmath}
\usepackage{amsfonts}
\usepackage{amssymb}
\usepackage{graphicx}
\usepackage{tabularx}
\usepackage{array}
\usepackage{cite}
\makeatletter
\def\slash#1{{\mathpalette\c@ncel{#1}}} 
\preprint{Cavendish--HEP--09/22}
\title{Distributions of charged massive scalars and fermions from evaporating higher-dimensional black holes}

\author{Marco O.P.\ Sampaio\\ 
  Cavendish Laboratory, University of Cambridge,
  J.J.\ Thomson Avenue, Cambridge CB3 0HE, U.K.\\ E-mail: \email{sampaio@hep.phy.cam.ac.uk}}

\abstract{A detailed numerical analysis is performed to obtain the Hawking spectrum for charged, massive brane scalars and fermions on the approximate background of a brane charged rotating higher-dimensional black hole constructed in~\cite{Sampaio:2009ra}. We formulate the problem in terms of a ``spinor-like'' first order system of differential wave equations not only for fermions, but for scalars as well and integrate it numerically. Flux spectra are presented for non-zero mass, charge and rotation, confirming and extending previous results based on analytic approximations. In particular we describe an inverted charge splitting at low energies, which is not present in four or five dimensions and increases with the number of extra dimensions. This provides another signature of the evaporation of higher-dimensional black holes in TeV scale gravity scenarios.}

\keywords{Black Holes, Hawking Radiation, Large Extra Dimensions}

\begin{document}

\section{Introduction}

In a recent paper~\cite{Sampaio:2009ra} an approximate background to model the gravitational field of a higher dimensional rotating black hole with a brane confined abelian charge was constructed. This background was coupled to massive, charged brane scalar and fermionic perturbations and the corresponding wave equations were separated to allow for the study of the associated Hawking radiation~\cite{Hawking:1974sw}. 

The detailed study of black hole perturbations is important in many different contexts such as astrophysics~\cite{Regge:1957td,Teukolsky:1973ha,Press:1973zz,Teukolsky:1974yv,Dolan:2008kf}, cosmology~\cite{Page:1976wx} or from the purely quantum field theoretical point of view~\cite{Hawking:2005kf,Rovelli:2008zza,Green:1987sp}. The particular construction in~\cite{Sampaio:2009ra} was motivated by TeV gravity scenarios which contain extra dimensions~\cite{Antoniadis:1990ew,Arkani-Hamed:1998rs,Antoniadis:1998ig,Arkani-Hamed:1998nn,Randall:1999vf,Randall:1999ee}, and the Standard Model fields confined to a \mbox{4-dimensional} brane~\cite{Flacke:2005hb,ArkaniHamed:1999dc,ArkaniHamed:1999za}. In such scenarios black holes may form~\cite{Giddings:2001bu,Dimopoulos:2001hw,Argyres:1998qn,Eardley:2002re} in the high energy collision of charged brane degrees of freedom so the brane charge of the black hole becomes important. In particular, recently, it has been shown numerically in four dimensions that black holes indeed form in ultra-relativistic collisions between Schwarzchild black holes~\cite{Sperhake:2008ga,Shibata:2008rq,Sperhake:2009jz} or between solitons~\cite{Choptuik:2009ww}. These results  support the early hoop conjecture type arguments.

In this paper we complete the study in~\cite{Sampaio:2009ra} by performing a full numerical analysis which is not constrained by the approximations of small energy, mass, charge and rotation used in analytic approaches. This allows us to obtain exact numerical results in the full range of energies which is important even if some of the parameters such as charge, mass or rotation are small. We confirm all the low energy features found in the analytic study, in particular the weakness of discharge for typical QED like charges. We also present plots for larger masses which are relevant for typical TeV gravity scenarios at the LHC.

The structure of the paper is the following: In Sec.~\ref{sec:fields} we briefly review the background fields and the wave equations for the scalar and fermionic perturbations with mass and charge. We formulate the scalar equation in a new ``spinor-like'' form which is particularly convenient for numerical integration. In Sect.~\ref{sec:numerical_methods} we provide a series expansions for the fields near the horizon, an asymptotic expansion at infinity and re-formulate the first order system in a more convenient form to extract the transmission factor. In Sect.~\ref{sec:numerical_results} we present plots for transmission factors and the number flux spectra, confirm the low energy results, extend them to the full energy range and point out the main features with and without rotation. Finally in Sec.~\ref{sec:Conclusions} we summarize the results emphasising the importance of implementing the new effects in black hole event generators~\cite{Dimopoulos:2001hw,Harris:2003db,Cavaglia:2006uk,Dai:2007ki,Frost:2009cf} to perform phenomenological studies of TeV gravity with black hole production.

\section{The fields}\label{sec:fields}

We want to study a charged rotating black hole background where the abelian (i.e. Maxwell) field is confined to a four dimensional brane. As derived in~\cite{Sampaio:2009ra} an approximate effective background in $4+n$ dimensions is characterized by the metric
\begin{multline}\label{ansatz_metric}
ds_{(4)}^2=\left(1-\dfrac{\bar{\mu} r^{1-n}-Q^2}{\Sigma}\right) dt^2+\dfrac{2a (\bar{\mu} r^{1-n}-Q^2) \sin^2{\theta}}{\Sigma}dt d\phi-\dfrac{\Sigma}{\Delta}dr^2-\\ -\Sigma d\theta^2-\left(r^2+a^2+\dfrac{a^2 (\bar{\mu} r^{1-n}-Q^2) \sin^2{\theta}}{\Sigma }\right)\sin^2{\theta}d\phi^2 \ , \end{multline} 
where
\begin{equation}
\Delta=r^2+a^2+Q^2-\dfrac{\bar{\mu}}{r^{n-1}}\,, \hspace{1cm} \Sigma=r^2+a^2\cos^2{\theta} \ ,
\end{equation}
$\bar{\mu}=1+a^2+Q^2$, if we take horizon radius units ($r_H=1$ -- see Sect.~4.1 of~\cite{Sampaio:2009ra}); and the Maxwell field is
\begin{equation}\label{KN_ansatz}
A_a dx^a = -Q \dfrac{r}{\Sigma}\left(dt-a\sin^2\theta d\phi\right) \ .
\end{equation} Here $Q$ is the electric charge of the black hole and $a$ its oblateness parameter which is directly related to the amount of angular momentum~\cite{Myers:1986un}.
If now we consider coupling other quantum fields of various spins, Hawking radiation is present~\cite{Hawking:1974sw,Unruh:1974bw,Gibbons:1975kk,Candelas:1981zv,Ottewill:2000qh,Casals:2005kr} and the fields are thermally emitted from the hole which evaporates progressively~\cite{Ida:2002ez,Harris:2003eg,Harris:2005jx,Ida:2005ax,Duffy:2005ns,Casals:2005sa,Cardoso:2005vb,Cardoso:2005mh,Ida:2006tf,Casals:2006xp,Casals:2008pq}. The various fluxes of particle number $N$, energy $E$, angular momentum $J$ and charge $Q$, are given by
\begin{equation}
\frac{d^2 \left\{N,E,J,Q\right\}}{dt d\omega} = \frac{1}{2\pi} \sum_{j=|s|}^\infty \sum_{m = -j}^{j} \frac{\left\{1,\omega,m,q\right\}}{\exp(\tilde{\omega}/T_H) \pm 1} \mathbb{T}^{(4+n)}_{k}(\omega,\mu, a,q,Q)\;,  \label{eq-fluxes}
\end{equation}
where $k=\left\{j,m\right\}$ are the angular momentum quantum numbers of the partial wave; $\omega,\mu,q$ are the energy, mass and charge of the particle respectively, $\tilde{\omega}=\omega-(ma+qQ)/(1+a^2)$ and 
\begin{equation}
T_{H}=\dfrac{(n+1)+(n-1)(a^2+Q^2)}{4\pi(1+a^2)r_H}
\end{equation}
is the Hawking temperature. The term containing the exponential in~\eqref{eq-fluxes} is the so called Planckian factor.
The transmission factor $\mathbb{T}^{(4+n)}$ is the fraction of a wave incident from infinity which is transmitted down the horizon and is purely ingoing at the horizon. This factor is obtained by solving the  wave equations for the particular field with such ingoing boundary conditions at the horizon. For the background~\eqref{ansatz_metric},~\eqref{KN_ansatz} separation of variables yields the following radial and angular equations~\cite{Sampaio:2009ra}:
\begin{itemize}
\item
{\em Massive charged scalars:}
\begin{equation}\label{eq:radial2s0}
\Delta\dfrac{d}{dr}\left(\Delta \dfrac{dR}{dr}\right)+\left(K^2-\Delta U\right)R=0 \; ,
\end{equation} 
where
\begin{eqnarray}\label{K_U_def}
K&=& \omega(r^2+a^2)-a m-qQr \\
U&=& \mu^2r^2+\Lambda_{c,j,m}+\omega^2a^2-2a\omega m \ .
\end{eqnarray}
The boundary condition at the horizon is~\cite{Bardeen:1972fi,Sampaio:2009ra}
\begin{equation}\label{eq:boundary2s0}
R= x^{-i\frac{K_\star}{\delta_0}}\left(1+\ldots\right)
\end{equation}
with $x=r-1$, $K_\star=\omega(1+a^2)-a m-qQ$, and $\delta_0=n+1+(n-1)(1+a^2+Q^2)$ is the leading order coefficient of the expansion of $\Delta$ in powers of $x$. The angular eigenvalue $\Lambda_{c,j,m}$ is determined from the angular equation
\begin{equation}\label{eq:2s0angular}
\dfrac{1}{\sin\theta}\dfrac{d}{d\theta}\left(\sin\theta\dfrac{dS}{d\theta}\right)+\left(c^2\cos^2\theta-\dfrac{m^2}{\sin^2\theta}+\Lambda_{c,j,m}\right)S=0
\end{equation}
with $c^2=a^2(\omega^2-\mu^2)$, by imposing regularity of the solution at $\cos\theta=\pm 1$. For $a=0$ we have the closed form $\Lambda_{0,j,m}=j(j+1)$.

Eq.~\eqref{eq:radial2s0}, can be written as a first order system of differential equations. This will be useful to perform the numerical integration using a method similar to that for fermions. Since there is no unique way of reducing the second order equation to a first order system, we take advantage of the extra freedom to construct a spinor-like object with a conserved Wronskian and, simultaneously, an asymptotic behaviour at infinity which gives the transmission factor straightforwardly. It is then possible to show that a convenient choice is
\begin{equation}
P_{\pm 0}= \dfrac{\Delta^{\frac{1}{2}}}{2}\left(k R\mp i \dfrac{dR}{dr}\right) \; , 
\end{equation} 
where in principle $k$ can be an arbitrary constant but we set it to the momentum of the partial wave $k=\sqrt{\omega^2-\mu^2}$.
So the second order equation~\eqref{eq:radial2s0} is replaced by the first order coupled system\footnote{Here $s$ is the spin which we leave arbitrary since the same type of equation will hold for fermions.}
\begin{equation}\label{eq:system_master}
\dfrac{d\mathbf{P}_s}{dr}=\mathbf{M}_s(r)\mathbf{P}_s
\end{equation}
where 
\begin{eqnarray}
\mathbf{M}_0(r)&=& \dfrac{\Delta^{\prime}}{2\Delta}\hat{\mathbf{\sigma}}_1-\dfrac{1}{2}\left(\dfrac{V}{k}-k\right)\hat{\mathbf{\sigma}}_2+\dfrac{i}{2}\left(\dfrac{V}{k}+k\right)\hat{\mathbf{\sigma}}_3\; ,
\end{eqnarray} 
\begin{equation}
V=\dfrac{K^2}{\Delta^2}-\dfrac{U}{\Delta} \ ,
\end{equation} and $\hat{\mathbf{\sigma}}_i$ are the Pauli matrices.
Now, using~\eqref{eq:system_master}, conservation of the Wronskian is easily checked: 
\begin{equation}\label{eq:wronskian}
\dfrac{d}{dr}\left(\mathbf{P}_s^\dagger\hat{\mathbf{\sigma}}_3\mathbf{P}_s\right)=\dfrac{d}{dr}\left(|P_{+|s|}|^2-|P_{-|s|}|^2\right)=0 \ .
\end{equation}
The choice $k=\sqrt{\omega^2-\mu^2}$ ensures that $P_{\pm 0}$ picks respectively the outgoing/incoming part of the wave at infinity (see Sect.~\ref{sec:ff_expansions}).
\item
{\em Massive charged fermions:}
For fermions, the radial equation obtained in~\cite{Sampaio:2009ra} is already in the form~\eqref{eq:system_master} with
\begin{eqnarray}
\mathbf{M}_{\frac{1}{2}}(r)
&=&\dfrac{\lambda}{\Delta^{\frac{1}{2}}}\hat{\mathbf{\sigma}}_1-\dfrac{\mu r}{\Delta^{\frac{1}{2}}}\hat{\mathbf{\sigma}}_2+i\dfrac{K}{\Delta}\hat{\mathbf{\sigma}}_3\; .
\end{eqnarray}
So $\mathbf{P}_{1/2}$ obeys the same Wronskian relation~\eqref{eq:wronskian} as does~\eqref{eq:system_master}. Again, the incoming solution at the horizon takes the form 
\begin{equation}
\mathbf{P}_{\frac{1}{2}}\sim x^{-i\frac{K_\star}{\delta_0}}\left(\mathbf{a}_0+\dots\right)
\end{equation}
with $\mathbf{a}_0$ a constant spinor. The angular eigenvalue is obtained from the system of angular equations
\begin{equation}\label{eq:2s1angular}
\left[\dfrac{d}{d\theta}+2s\left(a\omega\sin\theta-\dfrac{m}{\sin\theta}\right)+\frac{1}{2}\cot\theta\right]S_{-s}=\left(2s\lambda+a\mu\cos\theta\right)S_{s} \; , 
\end{equation}
where $s=\pm 1/2$. Once again, in general, the eigenvalues are obtained by imposing regularity of the solution at $\cos\theta=\pm 1$.  When $a=0$, $\lambda=j+1/2$ with $j$ a positive semi-integer.  
\item
{\em Electromagnetic and other perturbations:}
Regarding electromagnetic perturbations, it is known in four dimensions that they couple to gravitational perturbations for the Kerr-Newman black hole (see for example Sect.~111, Chapter 11 of~\cite{Chandrasekhar:1985kt} and references therein). Similarly, in the higher dimensional case, we would expect them to couple to gravitational modes on the brane. However, in the limit of small charge the perturbations should approximately decouple. This is indeed the case and an approximation scheme was developed by Dudley and Finley~\cite{Dudley:1977zz}. It amounts to considering separately one perturbation (either electromagnetic or gravitational) while setting the other to zero on the fixed background. This approximation was used for example in~\cite{Kokkotas:1993ef} and~\cite{Berti:2005eb} to compute quasinormal modes. In~\cite{Berti:2005eb} this was compared to other methods to confirm the validity of the approximation for small $Q$ (the special case $J=0$ was used). The approximate second order wave equation for a perturbation of spin $s$ is~\cite{Berti:2005eb}
\begin{equation}\label{eq_2s_Dudley_Finley}
\Delta^{1-s}\dfrac{d}{dr}\left[\Delta^{1+s}\dfrac{dR}{dr}\right]+\left[K^2-is\dfrac{d\Delta}{dr}K+\Delta \left(2is\dfrac{dK}{dr}-\lambda\right)\right]R=0 \; ,
\end{equation} 
where $K$ is the same as in~\eqref{K_U_def}, but does not contain the particle charge $q$-term. This correctly reduces to the exact result for scalars and fermions when a background charge is present (if $K$ contains the $q$-term in~\eqref{K_U_def}) and it describes the electromagnetic or gravitational perturbations approximately, for small $Q$. 

An important feature of electromagnetic and gravitational perturbations, compared to scalars and fermions, is that they are electrically neutral. No electric coupling means that qualitatively, not much will change compared to the case of no background charge.  Specially for small charges we see from~\eqref{eq_2s_Dudley_Finley} that the charge of the background only enters through the $Q^2$ term in $\Delta$. This affects mostly the Hawking temperature which in the small $Q$ limit will simply rescale the flux curves without much difference in shape. In fact as noted in Fig.2 of~\cite{Sampaio:2009ra} for scalars and fermions, the effect of a small background charge on neutral particles is indeed small on the transmission factors and the flux curves are simply rescaled by the different Hawking temperature in the thermal factor. So qualitatively nothing changes for neutral scalars and fermions which obey~\eqref{eq_2s_Dudley_Finley} so we would expect the same for higher spins. Therefore we will not present numerical results for the electromagnetic or gravitational field, since they reduce to well studied cases (see e.g.~\cite{Harris:2003eg,Casals:2005sa,Ida:2006tf}) both qualitatively and in terms of implementation (the following constant shift $a^2\rightarrow a^2+Q^2$ in $\Delta$ is sufficient).  

Furthermore, if we assume electroweak symmetry is not restored outside the black hole and that the electrically charged weak vector boson $W$ and the neutral $Z$ provide a good effective description of the weak degrees of freedom, it is tempting to guess that~\eqref{eq_2s_Dudley_Finley} holds similarly for those perturbations (with $K$ containing the electric $q$-coupling). This is because the black hole background can only be electrically charged (or colour charged) so the backrgound values of the weak field perturbations vanish. Then we would expect~\eqref{eq_2s_Dudley_Finley} to be exact since there is no reason for the weak field perturbations to couple to the linearized gravitational perturbations. This is in contrast with the equations for electromagnetic perturbations where terms linear in the gravitational perturbations arise from linearising bilinears in the gravitational/electromagnetic fields around their background values. For weak $W$ and $Z$ field perturbations (as for scalars and fermions) such gravitational terms can not be present because even if they exist before linearisation, when evaluated on the background for the $W$ and $Z$ fields they are identically zero. 

Furthermore, because the $W$ and $Z$ fields are massive, they are described by a complex or a real Proca field respectively, which is an extra complication. 

Alternatively, if electroweak symmetry is restored in the region outside the black hole\footnote{This should be the case if the black hole size is smaller than the electroweak breaking scale which is typically $1/m_W$, the inverse mass of the $W$.}, then we have to use the fundamental weak gauge fields associated with the $SU(2)_L\times U(1)_Y$ sector of the Standard Model (instead of the electromagnetic, the $W$ and the $Z$ fields). 

Due to these extra complications, the detailed study of other vector perturbations will be treated elsewhere.
\end{itemize}

\section{Numerical methods}\label{sec:numerical_methods}
In this section we present the methods used to reduce the linear systems of equations at hand to initial value problems which are more convenient for numerical integration.
\subsection{Near horizon expansions}
The boundary condition at the horizon is most easily implemented through a series expansion. This allows for a high precision initialisation of the radial functions slightly away from the horizon to avoid numerical difficulties associated with the coordinate singularity.

The expansions we need are 
\begin{eqnarray}\label{eq:nh_expansions}
R&=&\displaystyle x^\alpha\sum_{m=0}^{+\infty}\alpha_m x^m \nonumber\\
\mathbf{P}_{\frac{1}{2}}&=&\displaystyle x^{\alpha}\sum_{m=0}^{+\infty}\mathbf{a}_m \left(\sqrt{x}\right)^m \; .
\end{eqnarray}
Note that $R$ can be used to initialise $\mathbf{P}_0$. By inserting into the wave equations~\eqref{eq:radial2s0} and~\eqref{eq:system_master} respectively we obtain the following recurrence relations
\begin{equation}\label{eq:recurrence}
\left\{
\begin{array}{rcll}
\alpha&=&\displaystyle - i\dfrac{K_\star}{\delta_0} & \vspace{2mm}\\
\alpha_0&=&1, \; \; \alpha_m=\displaystyle \dfrac{-1}{m(m+2\alpha)\delta_0^2}\left[(m+\alpha)\delta_0\bar{\gamma}_{m}+\sum_{k=0}^{m-1}\left(\gamma_{k}(k+\alpha)\delta_{m-k}+\alpha_k\sigma_{m-k}\right)\right] & \vspace{2mm}\\
\mathbf{a}_0&=&\displaystyle\left(\begin{array}{c} 0 \\ 1\end{array}\right), \; \; \; \;  \mathbf{a}_{m}=\displaystyle (\mathbf{N}_{0}-\delta_0(m+2\alpha))^{-1}\left[\mathbf{b}_m-\sum_{j=0}^{m-1}\mathbf{N}_{m-j}\mathbf{a}_{j}\right] & m\geq 1 \; \; .\vspace{2mm}
\end{array} \right.
\end{equation}
where a choice of normalisation was made, when setting $\alpha_0$ and $\mathbf{a}_0$. The various coefficients are defined in appendices~\ref{app:scalar_coeffs} and~\ref{app:fermion_coeffs}. Using expansions~\eqref{eq:nh_expansions} we have initialised $\mathbf{P}_s$ at $x=0.1$ by truncating the series at eighteenth order. A first estimate of the numerical error can be made by modifying this choice (we have used $x=0.05$ and $x=0.01$ as a check).\footnote{Throughout we have required an error $\varepsilon<10^{-4}$, for the transmission factors.}

\subsection{Far field expansions}\label{sec:ff_expansions}
Once the radial function is initialised, numerical integration routines can be used to propagate the solution away from the horizon according to~\eqref{eq:system_master}. When sufficiently away from the horizon, the transmission factor can be evaluated by comparing the numerically propagated solution with its asymptotic form at large $r$. An asymptotic expansion can be found in the form
\begin{equation}\label{eq:ff_asympt_series}
\mathbf{P}_s=e^{qr}r^{-\gamma}\sum_{m=0}^{+\infty}\mathbf{q}^s_mr^{-m} \; ,
\end{equation}
if we expand
\begin{equation}\label{eq:asymptM}
\mathbf{M}_{s}=\sum_{m=0}^{+\infty}\mathbf{M}^s_{m}r^{-m}
\end{equation}
and equate~\eqref{eq:system_master} order by order. The leading behaviour is
\begin{equation}
\mathbf{P}_s=Y_{s}^{(out)}e^{iy}y^{i\varphi}\mathbf{d}^{+}_s+Y_{s}^{(in)}e^{-iy}y^{-i\varphi}\mathbf{d}^{-}_s \; ,
\end{equation}
where $y=kr$, $Y_{s}^{(out)}$ and $Y_{s}^{(in)}$ are constants,
\begin{equation}
\varphi=\epsilon \frac{\omega}{k}-\sigma \frac{\mu}{k} \; ,
\end{equation}
\begin{equation}
\begin{array}{cc}
\displaystyle \epsilon=-qQ+\omega(1+a^2+Q^2)\delta_{n,0} \ \ \ \ \ \  & \displaystyle \sigma=\dfrac{\mu}{2}(1+a^2+Q^2)\delta_{n,0} \; ,
\end{array}
\end{equation}
and
\begin{equation}
\begin{array}{cccc}
\mathbf{d}^{+}_0=\left(\begin{array}{c}1 \\ 0\end{array}\right)  &\hspace{5mm} \mathbf{d}^{-}_0=\left(\begin{array}{c}0 \\ 1\end{array}\right) & \hspace{5mm} \mathbf{d}^{+}_{\frac{1}{2}}=\left(\begin{array}{c} 1 \\ -\frac{\mu}{\omega+k}\end{array}\right) & \hspace{5mm} \mathbf{d}^{+}_{\frac{1}{2}}=\left(\begin{array}{c}-\frac{\mu}{\omega+k} \\ 1\end{array}\right) \; .
\end{array}
\end{equation}
We can now factor out the dependence at infinity so that the leading asymptotic form for the upper(lower) component of the spinor becomes $Y_{s}^{(out)}$($Y_{s}^{(in)}$) respectively. This is achieved by performing a rotation on the spinor $\mathbf{P}_{s}$ such that it eliminates a fixed number of subleading terms in the asymptotic expansion~\eqref{eq:asymptM} (in practise we have eliminated the first two subleading terms). Then the new spinor $\mathbf{Q}_s$ is related to $\mathbf{P}_{s}$ through $\mathbf{Q}_{s}=\mathbf{R}_s\mathbf{P}_{s}$ and the the system to integrate becomes
\begin{equation}\label{eq:master_integrate}
\dfrac{d\mathbf{Q}_s}{dy}=\mathbf{A}_s\mathbf{Q}_s \; .
\end{equation}  
The explicit forms for $\mathbf{A}_s$ and $\mathbf{R}_s$ are given in appendix~\ref{app:new_matrices}. 

Finally, the transmission factor is computed from the definition by taking the limit
\begin{equation}\label{eq:Tratio}
\mathbb{T}_{s}=\lim_{r\rightarrow +\infty}\left(1-\left|\dfrac{Q_{+s}}{Q_{-s}}\right|^2\right)=1-\left|\dfrac{Y_{s}^{(out)}}{Y_{s}^{(in)}}\right|^2 \; 
\end{equation}
($\pm s$ for upper/lower component respectively) and an estimate of the error is obtained by varying the large $r$ used in the limit. Furthermore, with the normalisation chosen in~\eqref{eq:recurrence} we can evaluate the Wronskian~\eqref{eq:wronskian} at the horizon and use its conservation to obtain a second expression 
\begin{equation}\label{eq:Twronskian}
\mathbb{T}_{s}=\lim_{r\rightarrow +\infty}\dfrac{k W_{s}}{|Q_{-s}|^2}=\dfrac{k W_{s}}{|Y_{s}^{(in)}|^2} \; ,
\end{equation}
where 
\begin{equation}
W_0= K_\star \hspace{1cm} W_{\frac{1}{2}}=\dfrac{\omega+k}{2} \; .
\end{equation}
By comparing the results from~\eqref{eq:Tratio} and~\eqref{eq:Twronskian}, we obtain another estimate of the numerical errors.  Eq.~\eqref{eq:Twronskian} is particularly useful since it contains explicitly the zeros of the transmission factor in the numerator. 

To integrate~\eqref{eq:master_integrate}, a code was written in C++ using the Gnu Standard Library (GSL) numerical integration routines. This was checked against an independent code in Maple11.

\subsection{Angular eigenvalues and angular functions}
To determine the transmission factors when the rotation parameter $a$ is non-zero, it is necessary to solve the angular equations~\eqref{eq:2s0angular} and~\eqref{eq:2s1angular} numerically (no closed form is known for the angular eigenvalue when $a\neq 0$).

Using~\eqref{eq:2s0angular} and~\eqref{eq:2s1angular}, it is easy to show that all cases, except for massive fermions in a rotating background, are described by the following second order equation:
\begin{equation}\label{eq:master_angular_restricted}
\dfrac{1}{\sin\theta}\dfrac{d}{d\theta}\left(\sin\theta\dfrac{dS_s}{d\theta}\right)+\left(c^2\cos^2\theta-2sc\cos\theta-\dfrac{(m+s\cos\theta)^2}{\sin^2\theta}+\Lambda_{c,j,m}+s\right)S_s=0
\end{equation}
where for fermions $\Lambda_{c,j,m}=\lambda^2-a^2\omega^2+2a\omega m-|s|-s$. This equation, which describes spheroidal harmonics, has been studied extensively in the literature. Whenever we evaluate the result with $a \neq 0$, we adopt the method in Appendix~D of~\cite{Frost:2009cf} to obtain the angular eigenvalues.

For massive fermions on a rotating background, Eq.~\eqref{eq:master_angular_restricted} will contain an extra term linear in $a\mu \, dS_s/d\theta$ (see Eq.~(4.29) of~\cite{Sampaio:2009ra}) so the method of~\cite{Frost:2009cf} cannot be applied. Nevertheless, since we are mostly interested in studying mass and charge effects, Eq.~\eqref{eq:master_angular_restricted} allows us to obtain a representative set of cases\footnote{Note that except for some final plots (which are present for illustration purposes), all the results in Sect.~\ref{sec:numerical_results} will have $a=0$.}.

\section{Numerical results}\label{sec:numerical_results}
In this section, samples of numerical data of transmission factors were generated using the method presented in Sect.~\ref{sec:numerical_methods}. From such data all interesting fluxes and distributions can be computed quickly. Most of the samples were generated up to $\omega = 10$, but some up to $\omega =5$ to save computing time. We show plots with $\omega<5$ since the curves are very quickly stabilised for large $\omega$ (either to a constant or a suppressed tail). 

In sections~\ref{sec:tfactors} and~\ref{sec:fluxes} we focus mostly on results with the rotation parameter off and describe the main features for different charges and masses in the full energy range\footnote{The effect of rotation was studied previously in~\cite{Ida:2002ez,Harris:2005jx,Ida:2005ax,Duffy:2005ns,Casals:2005sa,Ida:2006tf,Casals:2006xp,Casals:2008pq}, and massive fermions without rotation and at low energies in~\cite{Rogatko:2009jp}. Here we are mainly interested in charge and particle mass.}. For illustration purposes, in section~\ref{sec:fluxes} we present some curves with typical rotation and typical charges which may be relevant for TeV gravity scenarios.

\subsection{Transmission factors}\label{sec:tfactors}
\begin{figure}[t]
\begin{center}
\includegraphics[scale=0.65,clip=true,trim=2cm 0cm 0cm 0cm]{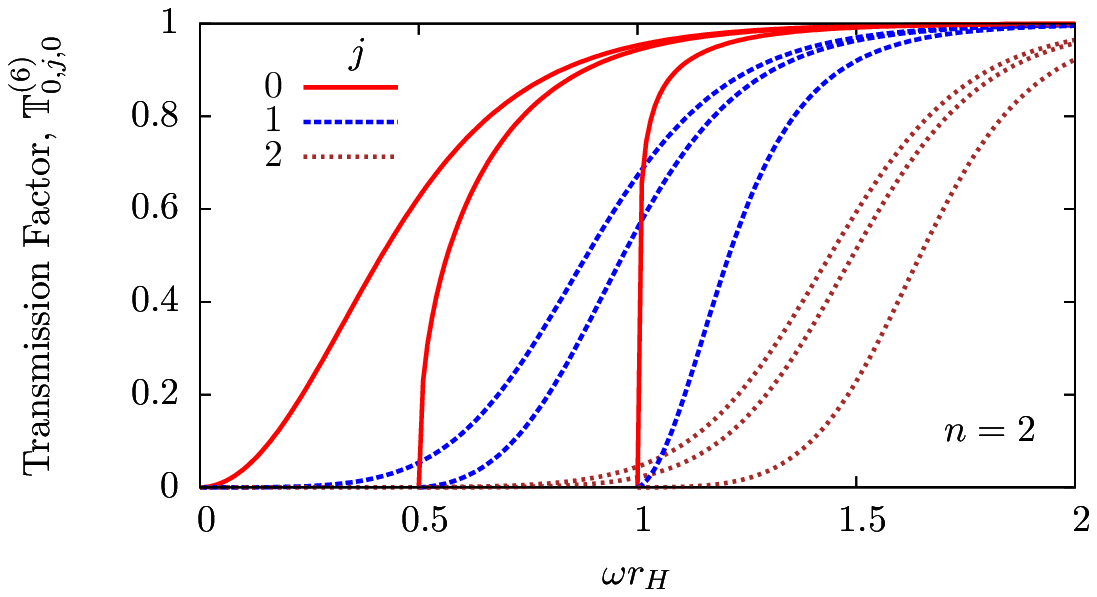}
\includegraphics[scale=0.65,clip=true,trim=2cm 0cm 0cm 0cm]{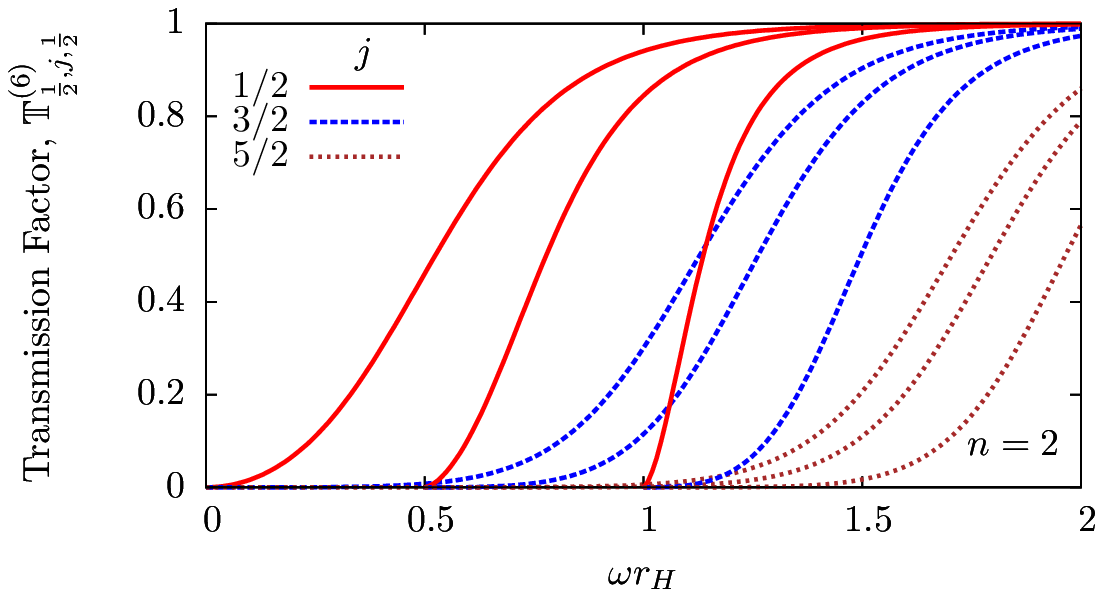}
\includegraphics[scale=0.65,clip=true,trim=2cm 0cm 0cm 0cm]{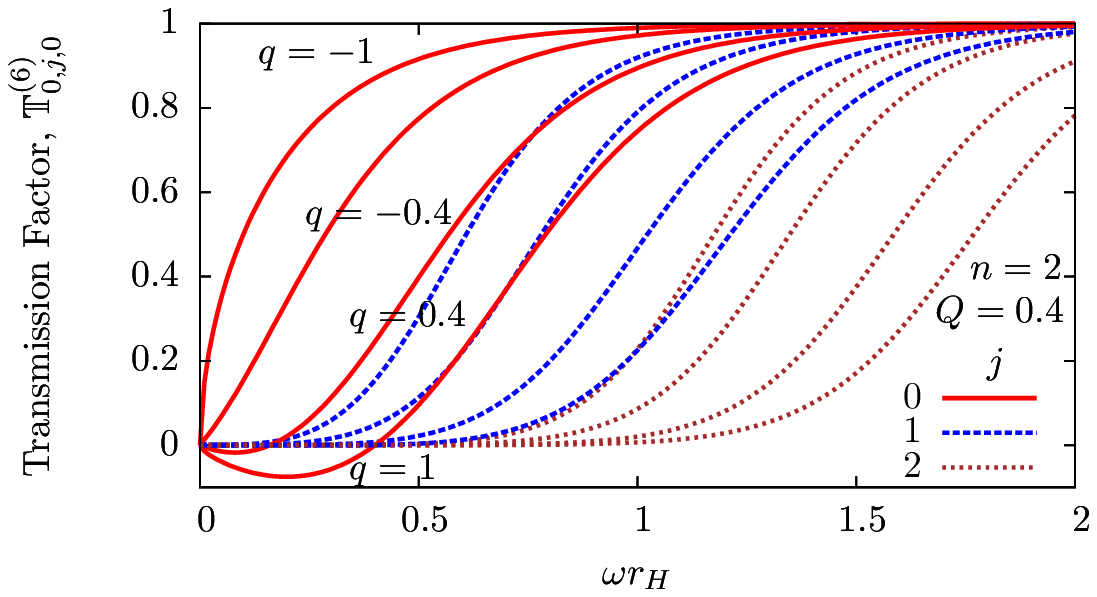} 
\includegraphics[scale=0.65,clip=true,trim=2cm 0cm 0cm 0cm]{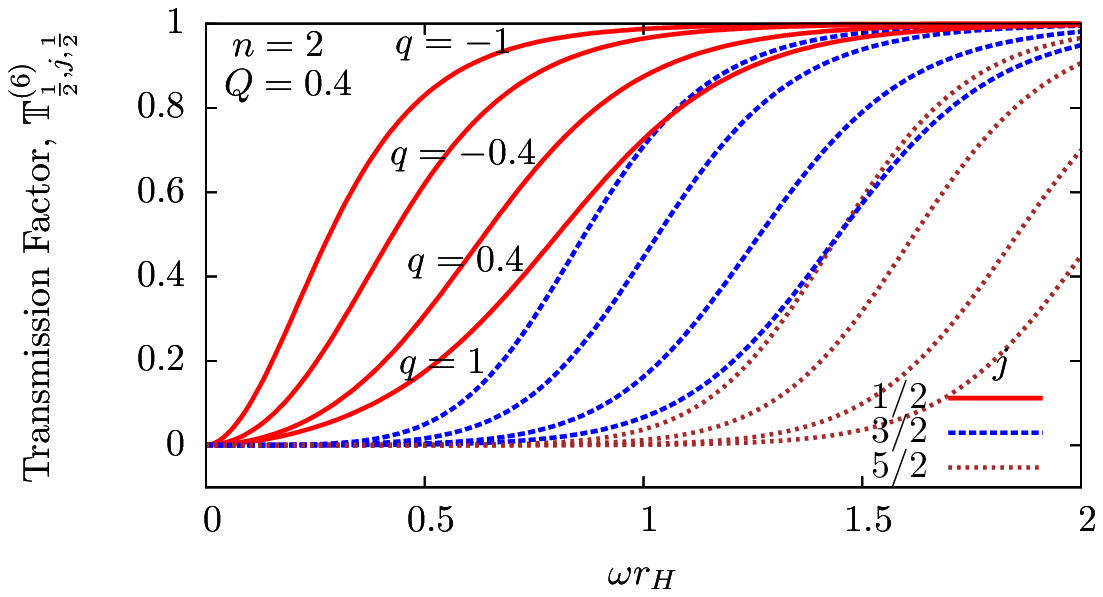}
\end{center}
\caption{\emph{Scalar (left) and fermion (right) transmission factors for $n = 2$}: The top plots are for variable $\mu$ and the bottom plots are for variable $q$. The first three partial waves are presented (note that $a=0$, so waves with different $m$ for the same $j$ are degenerate). \label{fig:tfactors_nfixed}}
\end{figure}
In this section we display transmission factors for individual modes, confirm the earlier results obtained in~\cite{Sampaio:2009ra} and extend them to the full energy range. We describe the main features of the plots which are relevant to the discussion of the fluxes in the next sections. We have checked that the approximate results based on the analytic approximations of~\cite{Sampaio:2009ra} reproduce well the exact numerical results obtained with our method even at intermediate energies as claimed there. 

In Fig.~\ref{fig:tfactors_nfixed} we present plots for the $n=2$ case and a range of charges and masses, with $\omega \in [0,2]$.\footnote{The transmission factors asymptote to unity quickly, so this is the interesting region.}
The top plots show the first three partial waves for scalars and fermions and $\mu=0$, $0.5$ and $1$. For scalars we confirm the strong suppression at the mass threshold for the $j=0$ partial waves, and the shift and suppression for higher partial waves~\cite{Sampaio:2009ra}. For fermions the behaviour is similar, except that the first partial waves are not so sharply suppressed at threshold. 
The bottom plots show the same partial waves when the mass is set to zero, the background charge is set to $Q=0.4$ and the particle charge varies between $q=1$ and $q=-1$. Again we confirm, for both scalars and fermions, that negative charges are favoured in the full range of energies (all curves split following the same pattern as indicated for the first mode). This is because the transmission factor is the fraction of a wave incident from infinity that is transmitted through the horizon, so we would expect the Coulomb attraction to favour such negative charges as observed. This feature will be important to understand the behaviour of the fluxes, in the next section. 

\begin{figure}[t]
\begin{center}
\includegraphics[scale=0.65,clip=true,trim=2cm 0cm 0cm 0cm]{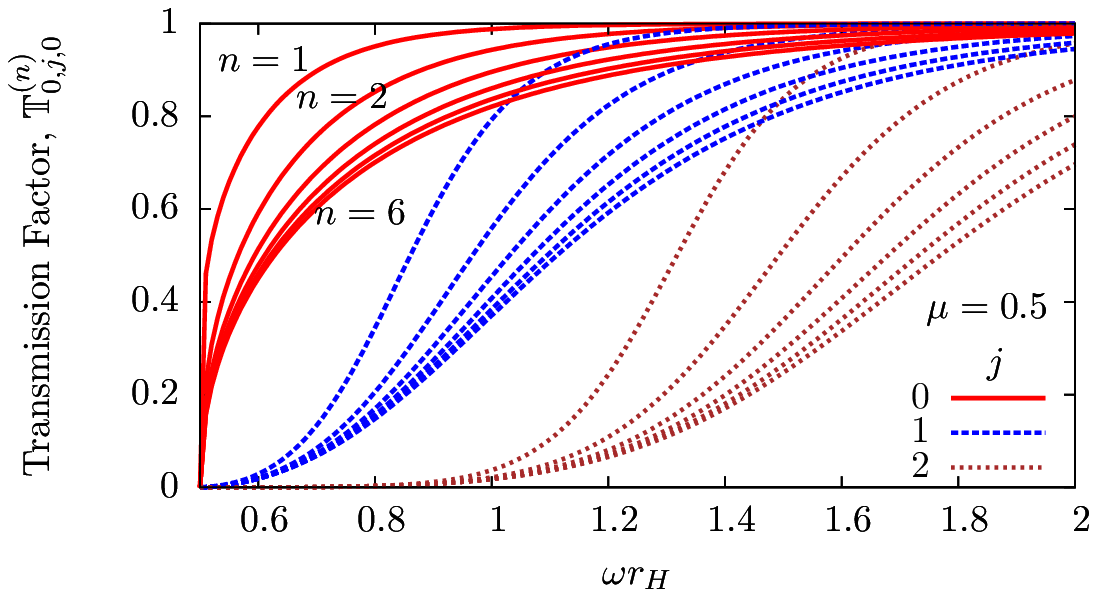}
\includegraphics[scale=0.65,clip=true,trim=2cm 0cm 0cm 0cm]{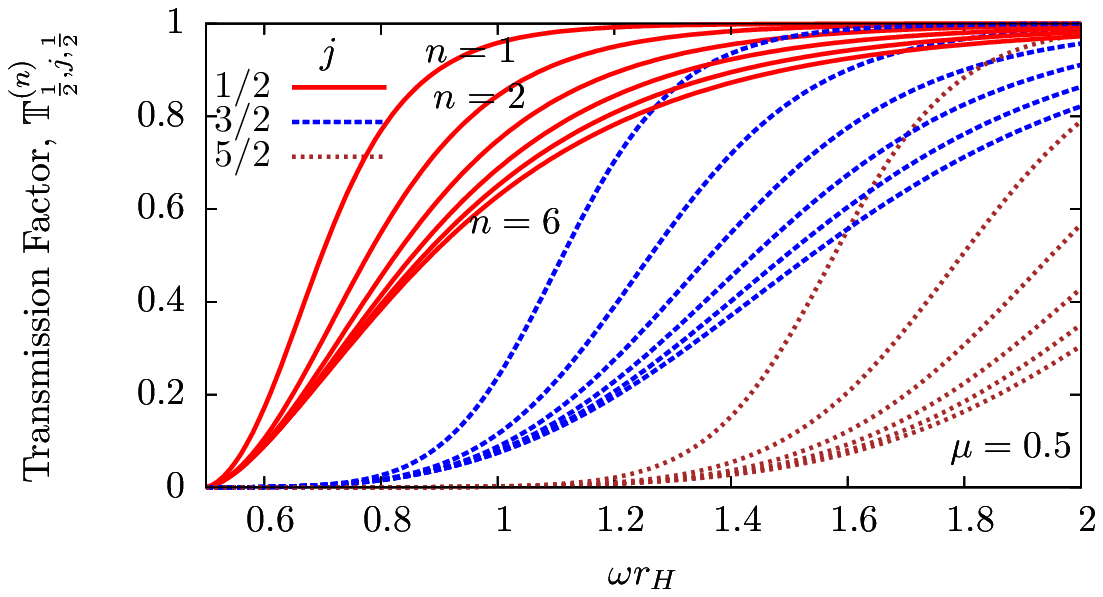}
\includegraphics[scale=0.65,clip=true,trim=2cm 0cm 0cm 0cm]{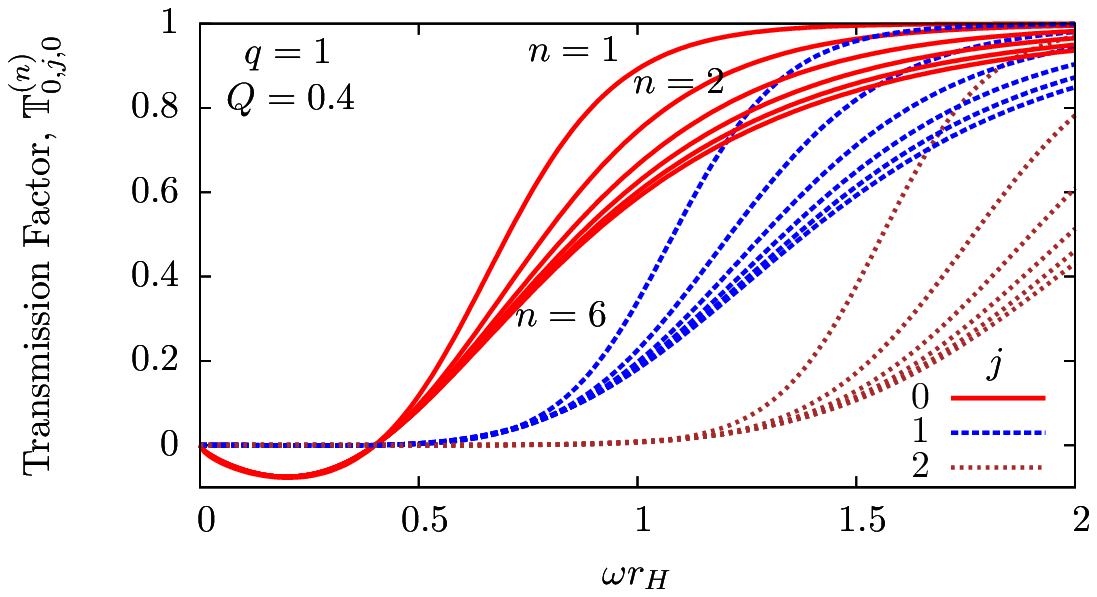} 
\includegraphics[scale=0.65,clip=true,trim=2cm 0cm 0cm 0cm]{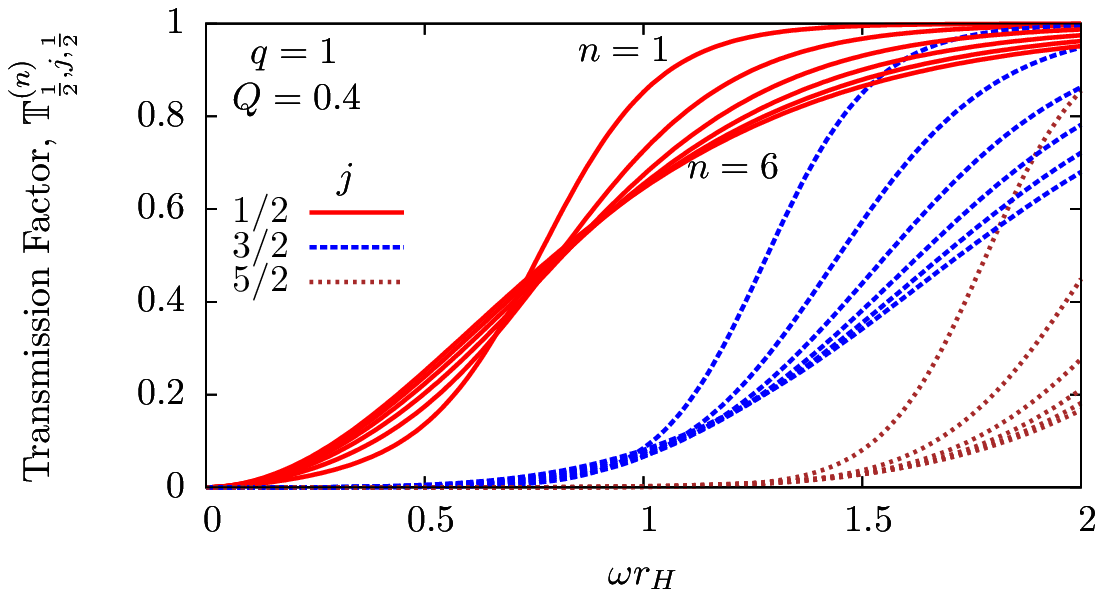}
\end{center}
\caption{\emph{Scalar (left) and fermion (right) transmission factors for variable $n$}: The top plots are for variable $\mu$ and the bottom plots are for variable $q$. Within partial waves with the same $j$, the curves are ordered from $n=1$ to $n=6$ from top to bottom (e.g. $j=1$ curves indicated in the plots).
\label{fig:tfactors_nVar}}
\end{figure}

Figure~\ref{fig:tfactors_nVar} shows the variation with $n$. The top plots are for $\mu=0.5$ and the bottom plots for $q=1$ and $Q=0.4$. The general tendency is for the transmission factor to be suppressed with $n$. The exception is in the low energy region, when the charge is non-zero, where the tendency is inverted for fermions, whereas for scalars in the superradiant region the variation with $n$ is small (this agrees with the results in~\cite{Sampaio:2009ra}).

\subsection{Fluxes}\label{sec:fluxes}

In this section we present plots for the particle number flux in the full energy range, summed over partial waves (we have included the first ten $j$-partial waves). This quantity is sufficient to illustrate the new effects, since for example the power flux curves are qualitatively similar and the charge fluxes are proportional to the corresponding number flux. Furthermore, the angular momentum flux (when rotation is present) has been studied in detail before~\cite{Ida:2002ez,Harris:2005jx,Ida:2005ax,Duffy:2005ns,Casals:2005sa,Ida:2006tf,Casals:2006xp,Casals:2008pq} and we will observe that even when the new effects are present, the contribution from rotation affects the spectrum in a similar fashion as in those earlier studies. 

\begin{figure}[t]
\begin{center}
\includegraphics[scale=0.65,clip=true,trim=1.8cm 0cm 0.1cm 0cm]{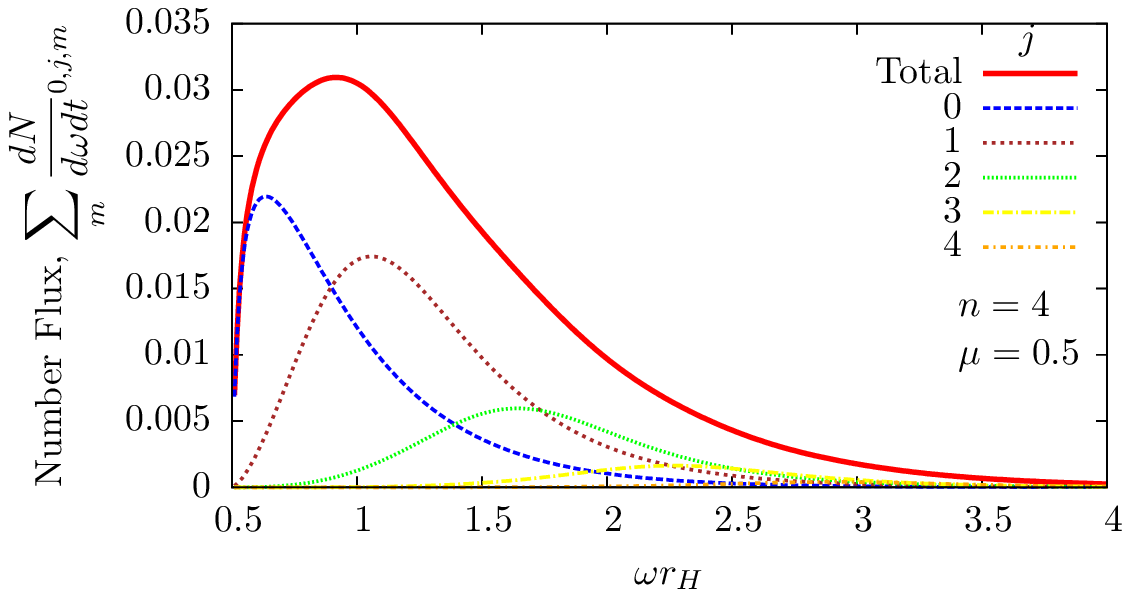}
\includegraphics[scale=0.65,clip=true,trim=1.6cm 0cm 0.1cm 0cm]{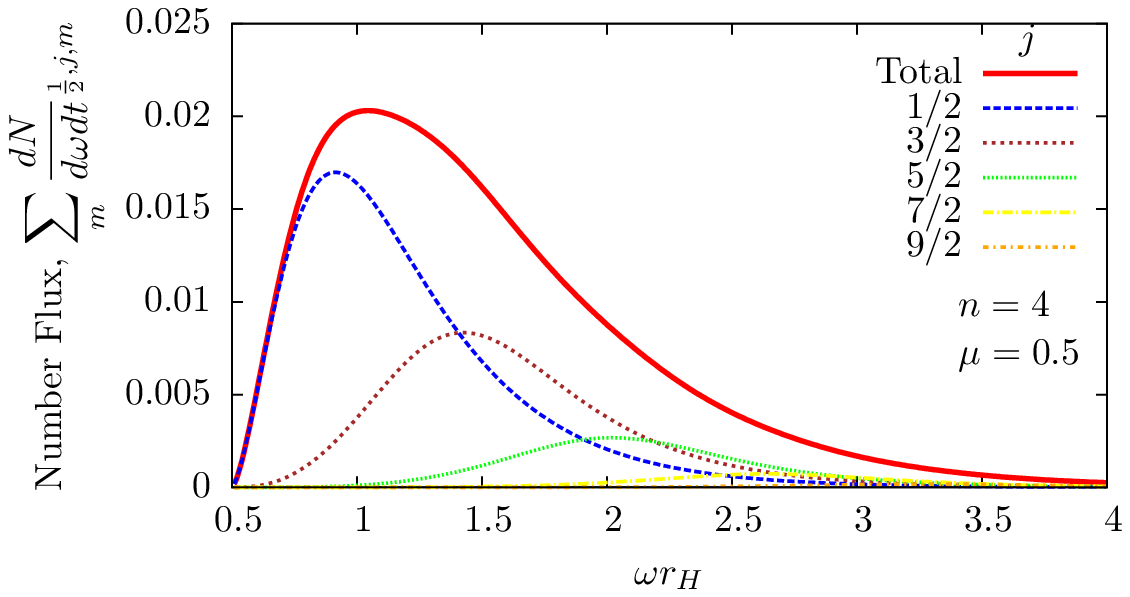} 
\includegraphics[scale=0.65,clip=true,trim=1.2cm 0cm 0.1cm 0cm]{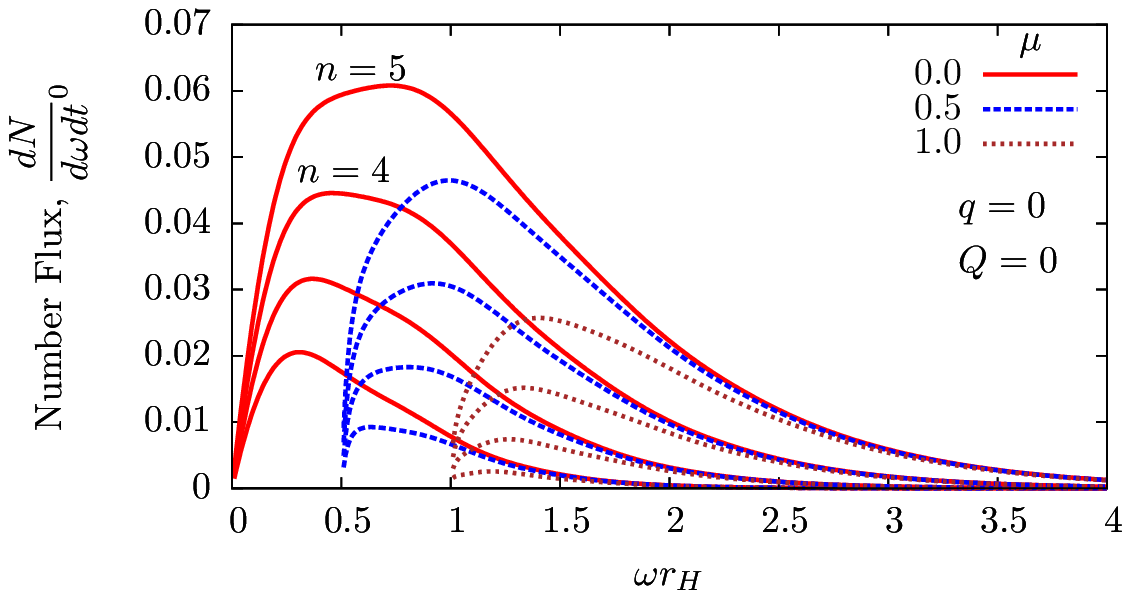}
\includegraphics[scale=0.65,clip=true,trim=1.2cm 0cm 0.1cm 0cm]{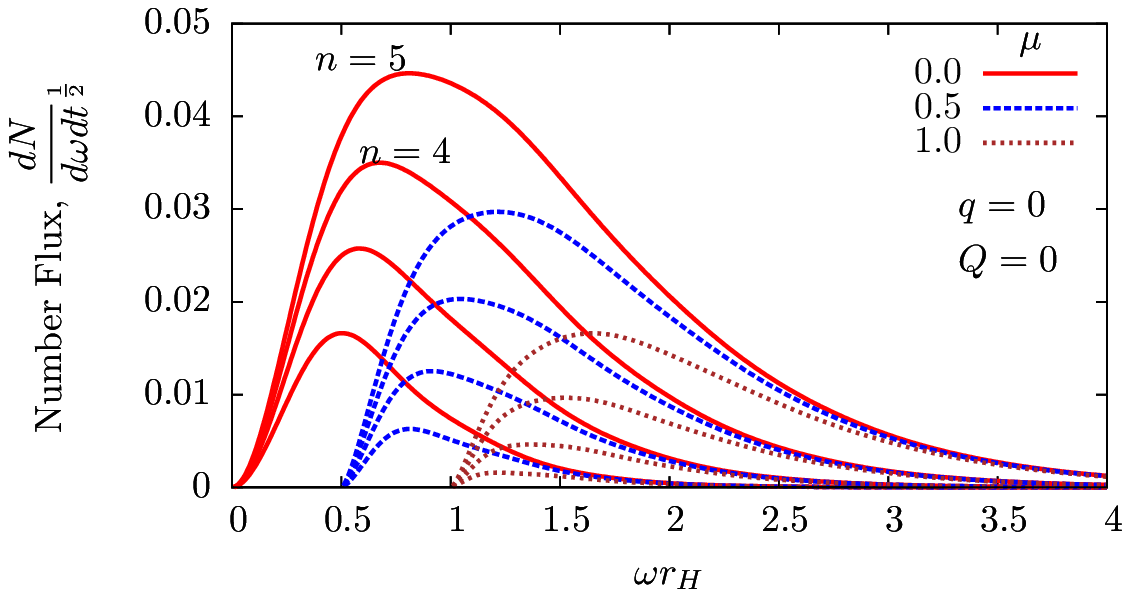}
\end{center}
\caption{\emph{Scalar (left) and fermion (right) number fluxes for $n = 4$ and zero charges}: The top plots show $\mu=0.5$ and the contributions from each partial wave to the total flux. The bottom plots show variable $\mu$ and variable $n=5,\ldots,2$. For each $\mu$ the curves are naturally order in $n$ from top to bottom, $n=5$ and $n=4$ are indicated for $\mu=0$.}\label{fig:fluxes_mu}
\end{figure}

In Fig.~\ref{fig:fluxes_mu} we present examples of the number flux for non-zero mass, when the charge and the rotation are set to zero. The top plots show $n=4$ and $\mu=0.5$. Note that typical values of $\mu$, for Standard Model heavy particles such as the top quark in TeV gravity scenarios,\footnote{$\mu$ is in horizon radius units and $1/r_H$ is typically in the range $200~\mathrm{GeV}-1000~\mathrm{GeV}$ for TeV gravity scenarios at the LHC.} range from $0.1$ to $0.5$. We also indicate the contributions from the first few $j$ values to the total flux curve. Similarly to the transmission factors, the main feature is a sharp suppression at threshold. The area under the curves is larger for scalars than fermions, which agrees with earlier studies (see for example~\cite{Duffy:2005ns,Casals:2006xp}). The bottom plots show three values of the mass and various $n$ values. We confirm the conclusion of~\cite{Sampaio:2009ra} that the area under the curves is suppressed as $\mu$ increases and the suppression at threshold is smooth both for scalars and fermions. The error from using the $\mu=0$ curve with a sharp cut at the mass is therefore large (most notably for fermions). Regarding variation with $n$, it is opposite to the tendency for the transmission factors so the $n$ dependence of the Planckian factor dominates the magnitude.   

\begin{figure}[t]
\begin{center}
\includegraphics[scale=0.65,clip=true,trim=1.8cm 0cm 0cm 0cm]{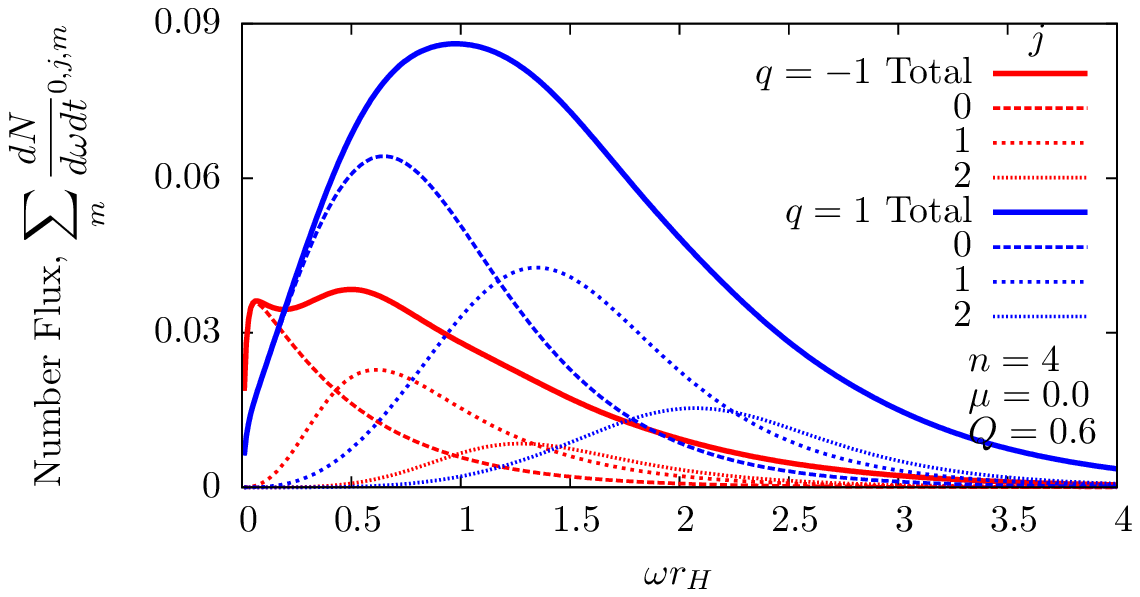}
\includegraphics[scale=0.65,clip=true,trim=1.9cm 0cm 0cm 0cm]{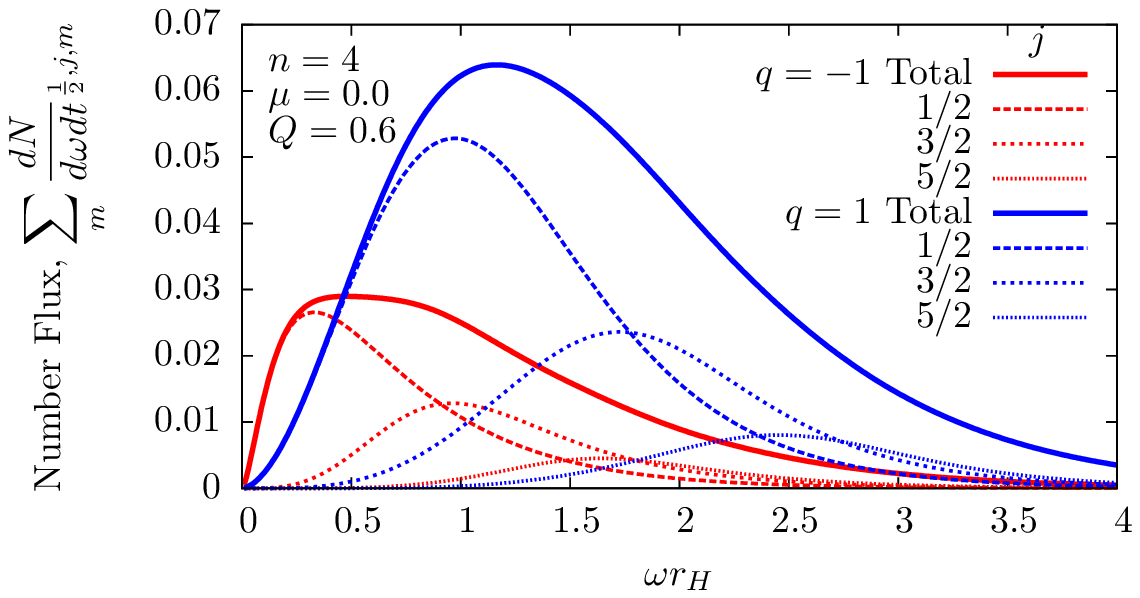}
\includegraphics[scale=0.65,clip=true,trim=1.4cm 0cm 0cm 0cm]{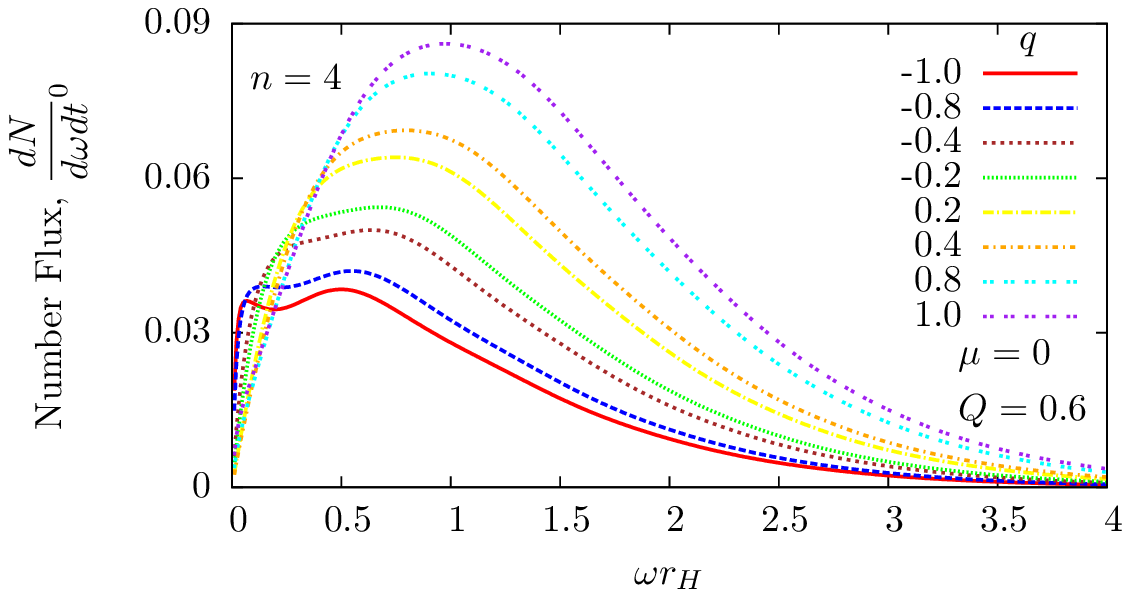} 
\includegraphics[scale=0.65,clip=true,trim=1.4cm 0cm 0cm 0cm]{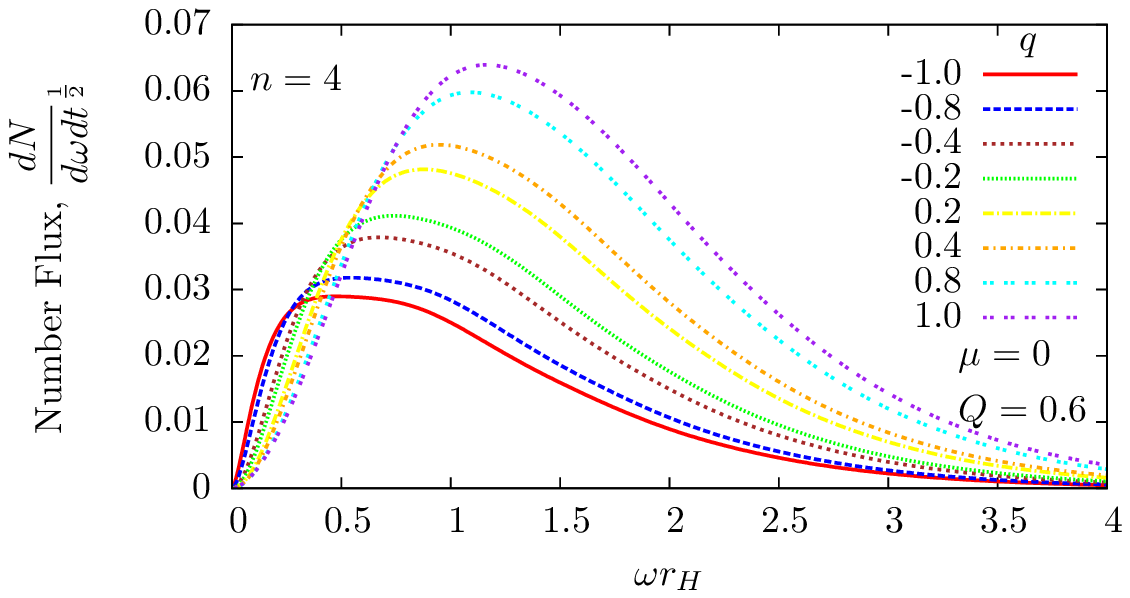}
\end{center}
\caption{\emph{Scalar (left) and fermion (right) number fluxes for $n = 4$, variable $q$, and $Q=0.6$}: The top plots show two opposite and large $|q|=1$ cases to illustrate the charge splitting, together with the first three partial wave contributions. The bottom plots show the variation of the curves between these two large charges.}
\label{fig:fluxes_charge}
\end{figure}

Figure~\ref{fig:fluxes_charge} shows several cases of non-zero charges. We have kept $Q=0.6$, which is a large value (see Sec.~3 of~\cite{Sampaio:2009ra}) so that all effects can be seen easily. Similarly we show $q$ in the range $[-1,1]$. We use $n=4$ as a representative case. The top plots show the total flux for the two extreme cases $q=1$ and $q=-1$ together with the first few partial waves contributing. The first striking observation is the confirmation that for all partial waves there is a region at low energy where charging up is favoured (i.e. the curve corresponding to negative charge is higher) and then another (dominant) region where discharge is favoured (the curve with positive charge is higher). It is also clear that if we integrate over the curves discharge is always favoured as expected. The bottom plots show a similar behaviour for a range of intermediate charges. Another interesting point is that the splitting at low energies is larger for fermions than for scalars. 

\begin{figure}[t]
\begin{center}
\includegraphics[scale=0.65,clip=true,trim=1.2cm 0cm 0.1cm 0cm]{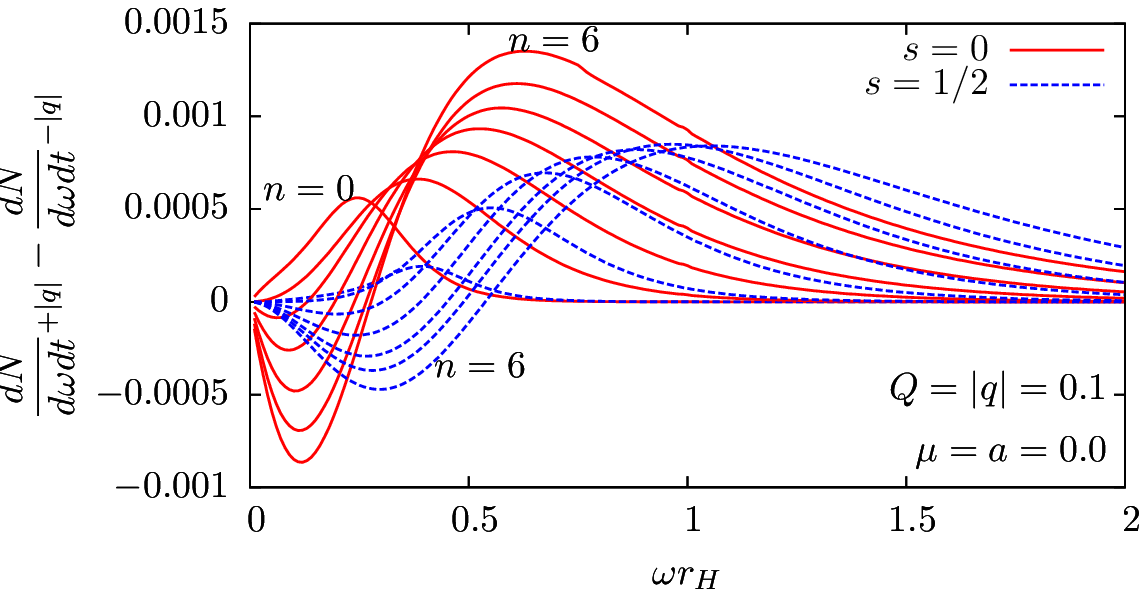}
\includegraphics[scale=0.65,clip=true,trim=1.2cm 0cm 0.1cm 0cm]{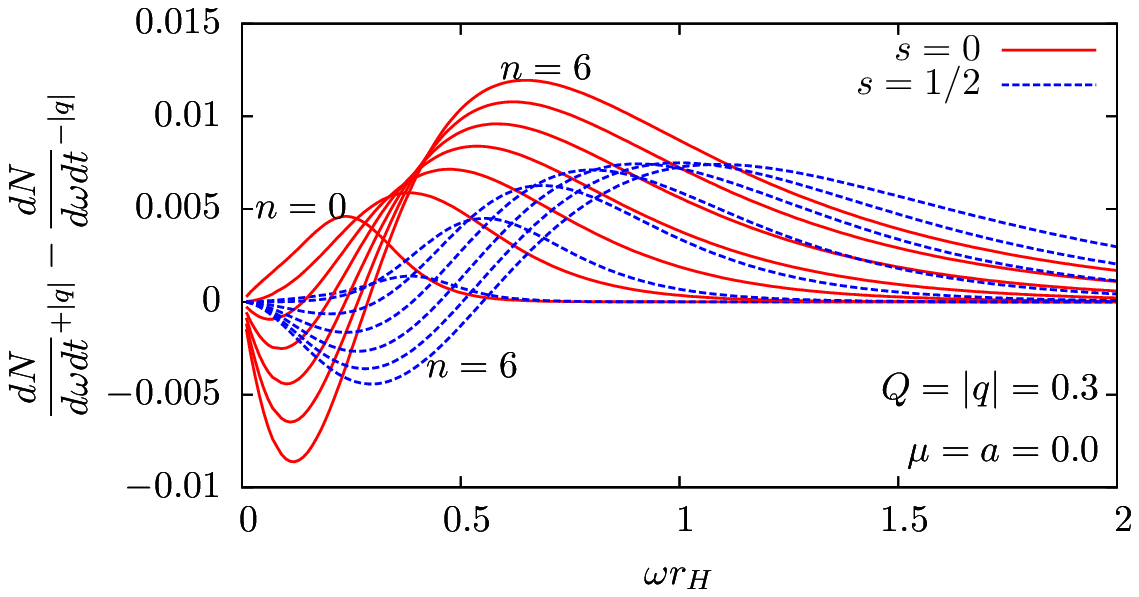}
\end{center}
\caption{\emph{Scalar and fermion number flux asymmetries}: Both plots show curves for the difference in number fluxes between positively charged and negatively charged particles for two values of $|q|Q$. The curves are naturally ordered in $n$ (some cases are labelled)  from $n=0$ (curve with the lowest maximum) to $n=6$ (highest maximum).}\label{fig:flux_split_a0}
\end{figure}

The inverted splitting at low energies is a direct consequence of the extra dimensions. In Fig.~\ref{fig:flux_split_a0} we show the difference in number flux of positively and negatively charged scalars and fermions when $n=0,\ldots,6$. The left plot shows a typical QED coupling of $|q|=Q=0.1$ and the right plot a QCD like coupling of $|q|=Q=0.3$. Note however that we are dealing with an abelian theory so the latter is only indicative of the magnitude of the effect for QCD. From this figure it is now clear that the splitting is controlled by an interplay between the transmission factor (which prefers negative charges) and the Planckian factor (which prefers positive charges). For $n=0$ and $n=1$, the splitting is always positive so the Planckian factor dominates. However as $n$ increases, the transmission factor starts dominating at low energies and for all $n\geq 2$ we have the observed inverted region (where the curves are negative). Another interesting feature of Fig.~\ref{fig:flux_split_a0} is that the plots on the left have exactly the same shape as the ones on the right. This is not surprising if we note  that for $qQ$ small we can expand the fluxes perturbatively around $qQ=0$ and since $|qQ|$ is $0.01$ and $0.09$ respectively, we would expect the perturbation to be dominated by the linear term so the difference is proportional to $|qQ|$.

 \begin{figure}[t]
\begin{center}
\includegraphics[scale=0.65,clip=true,trim=1.2cm 0cm 0cm 0cm]{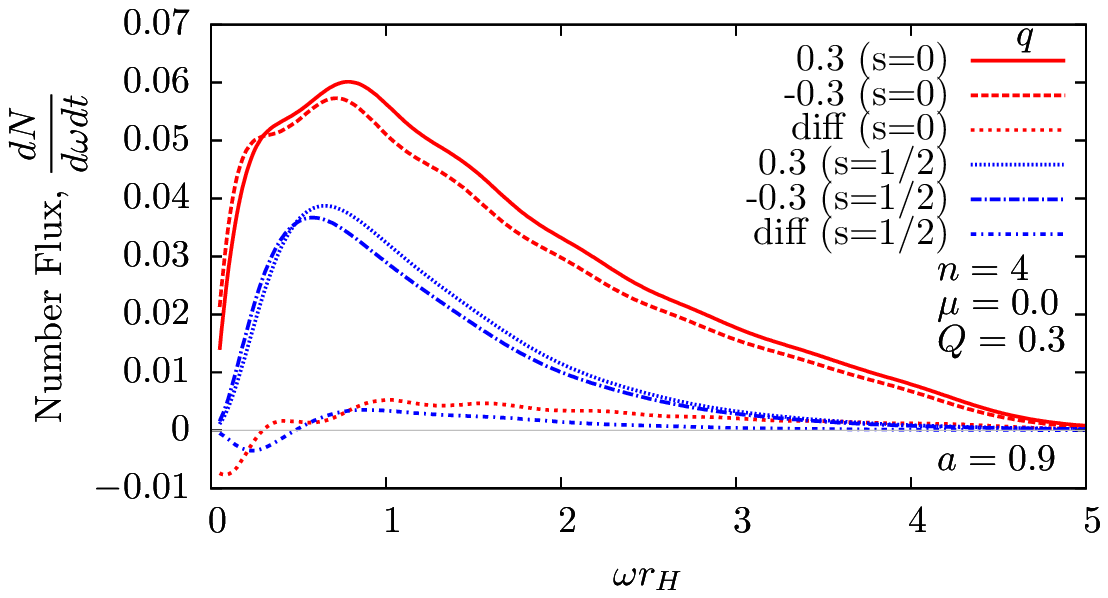}
\includegraphics[scale=0.65,clip=true,trim=1cm 0cm 0cm 0cm]{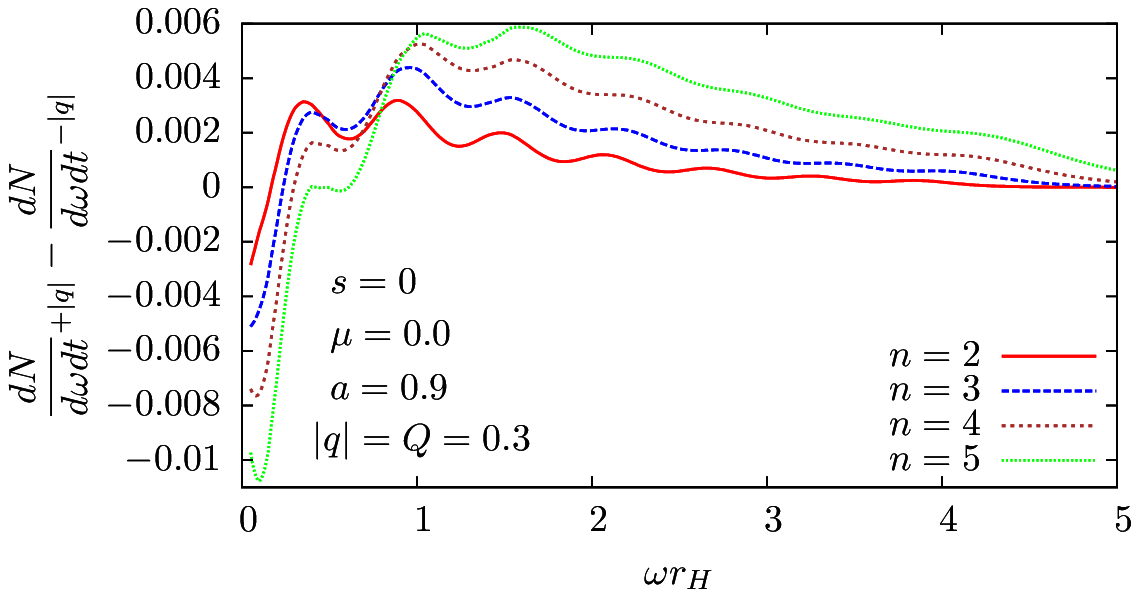}
\end{center}
\caption{\emph{Scalar and fermion asymmetries for $a=0.9$}: The left plot shows number fluxes for scalars and fermions with positive and negative $|q|=0.3$ for $n=4$. The difference between positive and negative $|q|$ curves is also shown in the same plot. The right plot shows the difference for scalars and a range of $n$'s.}\label{fig:flux_split_a1}

\end{figure}

Finally Fig.~\ref{fig:flux_split_a1} shows some cases with a rotation parameter $a=0.9$ (the typical order of magnitude for a TeV gravity scenario rotating black hole) and the typical QCD charges $|q|=Q=0.3$. The left plot shows the split flux curves for the QCD case both for scalars and fermions and the difference between the two. The right plot shows the difference curves for scalars and a range of $n$'s. Qualitatively, the splitting of the curves when $a \neq 0$ follows the same pattern as Fig.~\ref{fig:flux_split_a0}. The main differences are: the oscillations, which are due to the contribution of higher partial waves when $a\neq 0$ (they are responsible for shedding the angular momentum of the black hole~\cite{Ida:2002ez,Harris:2005jx,Ida:2005ax,Duffy:2005ns,Casals:2005sa,Ida:2006tf,Casals:2006xp,Casals:2008pq}); and the shift of the spectrum towards higher energies, which is again a well known effect of rotation related to the contribution of partial waves with larger $j$. It is interesting to note that the oscillations persist for large $n$ in the right plot, which is not true for the flux plots where they tend to be smoother~\cite{Ida:2002ez,Harris:2005jx,Ida:2005ax,Duffy:2005ns,Casals:2005sa,Ida:2006tf,Casals:2006xp,Casals:2008pq}.

\section{Conclusions}\label{sec:Conclusions}

In this paper we have performed a full numerical analysis of Hawking radiation for massive, charged scalars and fermions on an effective higher dimensional rotating black hole background with charge.  

In the first part we have re-formulated the problem in a convenient form and obtained: i) a series expansion to implement the boundary condition at the horizon, Eq.~\eqref{eq:nh_expansions}, and the asymptotic form at infinity, Eq.~\eqref{eq:ff_asympt_series}; and ii)  a``spinor-like'' first order system of differential equations both for scalars and fermions, Eq.~\eqref{eq:master_integrate}. 

In the second part we have shown a selection of plots to illustrate each effect.
The main results are:
\begin{itemize}
\item
  We confirm and extend all the conclusions in~\cite{Sampaio:2009ra} regarding massive particles to the full energy range, with all contributing partial waves. In particular we obtain the case of massive brane fermions for $n>0$, which was not studied before in the full energy range. The main difference is that for fermions the suppression is not so sharp at the threshold energy. Since the typical mass parameters of heavy Standard Model particles for TeV black hole scenarios can go up to $\sim 0.5$, this is an important effect.
\item
  Regarding charges, we have confirmed the splitting of the fluxes between positive and negative charges in the full energy range and showed that discharge is always favoured. The most interesting feature we have found is the inverted charge splitting at low energies which, we have shown, is a new effect due to the extra dimensions for $n\geq 2$. So, even though electric discharge may be small in TeV gravity black hole events~\cite{Sampaio:2009ra}, this splitting will still be present and it may be possible to reconstruct it if such events occur in future experiments. For QCD charges, the splitting should be even larger but a non-abelian analysis will be necessary to determine which observables will display it. 

These conclusions remain qualitatively the same with rotation, which affects the fluxes in ways that were observed in earlier studies.
  
\end{itemize}

To summarize, the methods we have described can be used in the full energy range to implement an improved model of the Hawking evaporation in black hole event generators with non-zero masses and charge asymmetries.   


\section*{Acknowledgements}

I thank colleagues in the Cambridge SUSY Working Group for helpful
discussions and suggestions. This work was supported by Funda\c c\~ao para a Ci\^encia e Tecnologia (FCT) - Portugal, grant SFRH/BD/23052/2005.


\section*{Appendices}
\appendix
\section{Expansion coefficients}

\subsection{Scalars}\label{app:scalar_coeffs}
The expansion coefficients we need are defined by
\begin{equation}\label{eq:expansions2s0}
\begin{array}{rcl}
\Delta&=&\displaystyle x\sum_{m=0}^{+\infty}\delta_m x^m \vspace{2mm}\\
K^2-\Delta U &=&\displaystyle  \sum_{m=0}^{+\infty}\sigma_m x^m \vspace{2mm} \\
\bar{\gamma}_m&=&\displaystyle \sum_{k=0}^{m-1}(k+\alpha)\alpha_k\delta_{m-k} \vspace{2mm} \\
\gamma_m&=&(m+\alpha)\alpha_m\delta_{0}+\bar{\gamma}_m \; \; .
\end{array} 
\end{equation}
It can be shown then that
\begin{eqnarray}
&\delta_0&=n+1+(n-1)\left(a^2+Q^2\right)  \\
&\delta_1&=1-\dfrac{n(n-1)\left(1+a^2+Q^2\right)}{2}  \\
&\delta_2&=\dfrac{n(n^2-1)\left(1+a^2+Q^2\right)}{6}  \\
&\delta_{m+1}&=-(1+\rho_{m+1})\delta_m \hspace{1cm} , m \geq 3
\end{eqnarray}
where
\begin{eqnarray}
\rho_2&=&\dfrac{n-2}{3}\\
\rho_{m+1}&=&\left(1-\dfrac{1}{m+2}\right)\rho_m
\end{eqnarray}
and
\begin{eqnarray}
&\sigma_0&=K_\star^2 \\
&\sigma_1&=2K_\star(2\omega-qQ)-U_0\delta_0  \\
&\sigma_2&=2K_\star\omega+(2\omega-qQ)^2-U_0\delta_1-U_1\delta_0  \\
&\sigma_3&=2\omega(2\omega-qQ)-U_0\delta_2-U_1\delta_1-U_2\delta_0  \\
&\sigma_4&=\omega^2-U_0\delta_3-U_1\delta_2-U_2\delta_1  \\
&\sigma_{m}&=-U_0\delta_{m-1}-U_1\delta_{m-2}-U_2\delta_{m-3} \hspace{1cm} , m \geq 5
\end{eqnarray}
where
\begin{eqnarray}
&U_0&=\Lambda+\omega^2a^2-2a\omega m+\mu^2 \\
&U_1&=2\mu^2 \\
&U_2&=\mu^2
\end{eqnarray}

\subsection{Fermions}\label{app:fermion_coeffs}
Similarly to the scalar case define
\begin{equation}
\begin{array}{rcl}
2\Delta \mathbf{M}_{\frac{1}{2}}(r)&=&\displaystyle \sum_{m=0}^{+\infty}\mathbf{N}_{m} \left(\sqrt{x}\right)^m \vspace{2mm} \\
\Delta^{\frac{1}{2}}&=& \displaystyle\sqrt{x}\sum_{m=0}^{+\infty}\bar{\delta}_m x^m \vspace{2mm} \\
\mathbf{b}_{2m}&=&\displaystyle \sum_{j=0}^{m-1}2\delta_{m-j}(j+\alpha)\mathbf{a}_{2j} \vspace{2mm} \\
\mathbf{b}_{2m+1}&=&\displaystyle \sum_{j=0}^{m-1}\delta_{m-j}(2j+2\alpha+1)\mathbf{a}_{2j+1} \; \; ,
\end{array} 
\end{equation}
The matrices we need are
\begin{eqnarray}
&\mathbf{N}_0&=2iK_\star \hat{\sigma}_3 \\
&\mathbf{N}_1&=2\lambda \bar{\delta}_0\hat{\sigma}_1-2\mu\bar{\delta}_0\hat{\sigma}_2\\
&\mathbf{N}_2&=2i(2\omega-qQ)\hat{\sigma}_3 \\
&\mathbf{N}_3&=2\lambda \bar{\delta}_1\hat{\sigma}_1-2\mu\left(\bar{\delta}_1+\bar{\delta}_0\right)\hat{\sigma}_2\\
&\mathbf{N}_4&=2i\omega\hat{\sigma}_3 \\
&\mathbf{N}_{2m}&=0 \; \; , m> 2\\
&\mathbf{N}_{2m+1}&=2\lambda \bar{\delta}_m\hat{\sigma}_1-2\mu(\bar{\delta}_m+\bar{\delta}_{m-1}) \hat{\sigma}_2 \; \; , m\geq 1
\end{eqnarray}
where $\bar{\delta}_i$ are obtained from the following expansion
\begin{eqnarray}
\Delta^{\frac{1}{2}}&=& =\sqrt{x}\delta_0^{\frac{1}{2}}\left(1+\sum_{m=1}^{+\infty}\dfrac{\delta_m}{\delta_0} x^m\right)^{\frac{1}{2}}  
\end{eqnarray}
by fixing a certain order of truncation and expanding the square root in powers of $x$ up to the given order.

\section{Matrices}\label{app:new_matrices}
In the main text we have used the following matrices:
\begin{equation}
\mathbf{R}_0=\left(\begin{array}{cc}& \vspace{-3mm}\\ e^{iy}y^{i\varphi}& \hspace{2mm}  0 \vspace{4mm}\\ 0  & e^{-iy}y^{-i\varphi} \vspace{2mm}\end{array}\right) \; 
\end{equation}

\begin{equation}
\mathbf{R}_{\frac{1}{2}}=\left(\begin{array}{cc}& \vspace{-3mm}\\e^{iy}y^{i\varphi} & \hspace{2mm}  0 \vspace{4mm}\\ 0  & e^{-iy}y^{-i\varphi} \vspace{2mm}\end{array}\right)\dfrac{1}{k(\omega+k)}\left(\begin{array}{cc}& \vspace{-3mm}\\\omega+k & \hspace{2mm}  -\mu  \vspace{4mm}\\ -\mu  & \omega+k \vspace{2mm}\end{array}\right) \; 
\end{equation}

\begin{equation}
\mathbf{A}_s=\left(\begin{array}{cc}& \vspace{-3mm}\\iB_s & \hspace{2mm}  \left(X_s+iY_s\right)e^{-i\Phi} \vspace{4mm}\\ \left(X_s-iY_s\right)e^{i\Phi}  & -iB_s\vspace{2mm}\end{array}\right) \; 
\end{equation}
with
\begin{equation}
\Phi =2\left(y+\varphi\log{y}-\sum_{m=1}^{j}\dfrac{c_m}{my^m}\right)
\end{equation}

\begin{equation}\label{eq:BsFactored}
B_s=\left\{\begin{array}{ll} \vspace{2mm}  \displaystyle \dfrac{V}{2k^2}-\frac{1}{2}-\dfrac{\varphi}{y}-\sum_{m=2}^{j}\dfrac{c_m}{y^m}&, s=0 \vspace{8mm} \\ \displaystyle \dfrac{\omega}{k^2}\dfrac{K}{\Delta}-\dfrac{\mu}{k^2}\dfrac{\mu r}{\Delta^{\frac{1}{2}}}-1-\dfrac{\varphi}{y}-\sum_{m=2}^{j}\dfrac{c_m}{y^m}&, s=1/2  \end{array}\right. \ .
\end{equation}
 $c_m$ are coefficients such that the corresponding powers in the asymptotic expansion of~\eqref{eq:BsFactored} are cancelled;
\begin{equation}
X_s=\left\{\begin{array}{ll} \vspace{2mm} \displaystyle \dfrac{1}{\Delta}\left(y+\dfrac{(n-1)\left(1+a^2+Q^2\right)k^{n+1}}{2y^n}\right)&, s=0 \vspace{8mm} \\\displaystyle \dfrac{\lambda}{\Delta^{\frac{1}{2}}}&, s=1/2  \end{array}\right.
\end{equation}
\begin{equation}
Y_s=\left\{\begin{array}{ll} \vspace{2mm} \displaystyle \dfrac{V}{2k^2}-\dfrac{1}{2}&, s=0 \vspace{8mm} \\\displaystyle \dfrac{1}{\Delta}\left(\dfrac{\mu \omega y}{k^2}\left(y-\Delta^{\frac{1}{2}}\right)+\omega\mu a^2-a\mu m-\dfrac{qQy}{k}\right)&, s=1/2  \end{array}\right.
\end{equation}
and now 
\begin{equation}
\Delta = y^2+k^2\left(a^2+Q^2\right)-\dfrac{\left(1+a^2+Q^2\right)k^{n+1}}{y^{n-1}} \ .
\end{equation}

\bibliography{ChargeNumeric}
\bibliographystyle{JHEP}

\end{document}